\documentclass[aps,prl,twocolumn,superscriptaddress,showpacs,amsfonts]{revtex4-1}
\usepackage{epsfig}
\usepackage{graphicx}
\usepackage{amsmath}
\usepackage{amssymb}
\usepackage{color}
\usepackage{bm}

\newcommand{\beq}{\begin{equation}}
\newcommand{\eeq}{\end{equation}}
\newcommand{\beqarray}{\begin{eqnarray}}
\newcommand{\eeqarray}{\end{eqnarray}}

\newcommand{\Hc}{\ensuremath{\mbox{H.c.}}} % Hermitian conjugate
\newcommand{\eq}[1]{Eq.~(\ref{#1})} % Eq. label
\newcommand{\fig}[1]{Fig.~\ref{#1}} % Fig. label
\allowdisplaybreaks

\begin{document}

\title{Spin-orbital coupling in a triplet superconductor--ferromagnet junction}

\author{Paola Gentile}
\affiliation{SPIN-CNR, I-84084 Fisciano (Salerno), Italy}
\affiliation{Dipartimento di Fisica ``E. R. Caianiello'', Universit\`{a} di
  Salerno, I-84084 Fisciano (Salerno), Italy }
\author{Mario Cuoco}
\affiliation{SPIN-CNR, I-84084 Fisciano (Salerno), Italy}
\affiliation{Dipartimento di Fisica ``E. R. Caianiello'',
Universit\`{a} di
  Salerno, I-84084 Fisciano (Salerno), Italy }
\author{Alfonso Romano}
\affiliation{SPIN-CNR, I-84084 Fisciano (Salerno), Italy}
\affiliation{Dipartimento di Fisica ``E. R. Caianiello'', Universit\`{a} di
  Salerno, I-84084 Fisciano (Salerno), Italy }
\author{Canio Noce}
\affiliation{SPIN-CNR, I-84084 Fisciano (Salerno), Italy}
\affiliation{Dipartimento di Fisica ``E. R. Caianiello'', Universit\`{a} di
  Salerno, I-84084 Fisciano (Salerno), Italy }
\author{Dirk Manske}
\affiliation{Max-Planck-Institut f\"{u}r Festk\"{o}rperforschung,
  Heisenbergstr. 1, D-70569 Stuttgart, Germany}
\author{P. M. R. Brydon}
\affiliation{Institut f\"{u}r Theoretische Physik, Technische Universit\"{a}t
  Dresden, D-01062 Dresden, Germany}

\date{\today}

\begin{abstract}
We study the interplay of spin and orbital degrees of freedom in a
triplet superconductor-ferromagnet junction. Using a
self-consistent spatially-dependent mean-field theory, we show
that increasing the angle between the ferromagnetic moment and the
triplet vector order parameter enhances or suppresses the $p$-wave
gap close to the interface, according as the gap antinodes are
parallel or perpendicular to the boundary, respectively. The
associated change in condensation energy establishes an
orbitally-dependent preferred orientation for the magnetization.
When both gap components are present, as in a chiral
superconductor, first-order transitions between different moment
orientations are observed as a function of the exchange field
strength.
%In this case the modification of the energy spectrum at
%the interface plays a relevant role.
\end{abstract}


\maketitle

\emph{Introduction.} The singlet superconductor (SSC) and
ferromagnet (FM) phases are fundamentally incompatible, as the
exchange field of the FM destroys the superconductivity by
aligning the anti-parallel spins of the electrons in singlet
Cooper pairs~\cite{SaintJames1969}. This pair-breaking effect
makes homogeneous coexistence of SSC and FM very rare. On the
other hand, SSC-FM interfaces can be readily fabricated in
artificial heterostructures, and the study of these devices has
attracted intense
attention~\cite{SCFMreviews,Ryazanov2001,Begeret2004,spinactive,crypto,Cuoco2008}.
The pair-breaking effect is central to the understanding of these
systems, e.g. it causes the spatial oscillation of the SSC
correlations in the barrier of a ferromagnetic Josephson junction,
which is responsible for the famed $0$-$\pi$
transition~\cite{SCFMreviews,Ryazanov2001}. The FM also suppresses
the SSC gap close to the
interface~\cite{SCFMreviews,spinactive,Cuoco2008}, and can induce
a magnetization in the SSC~\cite{Begeret2004}. Conversely, in
order to minimize pair-breaking in the SSC, the magnetization in
the FM may be suppressed near to the interface~\cite{Begeret2004},
while domains may spontaneously form in a thin FM
layer~\cite{crypto}.
% so as to give zero averaged magnetization on
%the scale of the coherence length

The coexistence of FM and triplet superconductor (TSC) states is
more favorable, as the exchange field is only pair breaking when
it is perpendicular to the Cooper pair spins. The physics of
TSC-FM devices is therefore richer than their singlet
counterparts, as the orientation of the FM moment relative to the
TSC vector order parameter is now a crucial variable. This is
predicted to control the nature of the proximity effect in TSC-FM
bilayers~\cite{Annunziata2011} and the sign of the current in
TSC-FM-TSC Josephson junctions~\cite{TFT}. In addition to the pair
breaking, spin-flip reflection processes at the interface with the
FM scatter the triplet Cooper pairs
%the interface with the FM
%also induces distinctly new physics within the
%TSC: spin-flip reflection processes
%scatter the triplet Cooper pairs
between the spin $\uparrow$ and
$\downarrow$ condensates, setting up a Josephson-like
coupling between them. The resulting ``spin Josephson effect'' is
manifested as a spontaneous spin current in the TSC normal to the
TSC-FM interface~\cite{Brydon2009,Brydon2011}.

The pair-breaking and spin Josephson coupling both make
significant contributions to the free energy of a TSC-FM junction
through the proximity effect, interface electronic reconstruction,
and the variation of the TSC gap. Although these contributions
depend upon the direction of the FM's exchange field, the two
effects do not necessarily act constructively: while pair-breaking
is always absent for a moment perpendicular to the TSC's vector
order parameter, the effective Josephson phase difference
%between the two spin condensates
can vanish for parallel and perpendicular configurations,
depending on the orbital pairing state. It is the purpose of this
Letter to explore in an unbiased way the
interplay of the spin- and orbital-structure of the TSC %state
in setting the stable orientation of the FM's moment. This
is a timely problem, as the recent
preparation~\cite{Krockenberger2010} of superconducting thin films
of Sr$_2$RuO$_4$~\cite{MacMae2003} opens the way to TSC
heterostructures. Furthermore, the proposed appearance of Majorana
fermions at TSC-FM interfaces in quantum wires motivates a deeper
understanding of the interplay between FM and
TSC~\cite{topological}.

To this purpose we study a lattice model of a TSC-FM
heterostructure using a self-consistent Bogoliubov-de Gennes
theory~\cite{Cuoco2008,Kuboki2004}. For a single-component
$p$-wave TSC, we find that the variation of the gap controls the
orientation of the FM's moment via the change in condensation
energy. The stable configuration is either
%The FM moment is stabilized either
parallel or perpendicular to the TSC vector order parameter,
depending on the alignment of the TSC gap with respect to the
interface, thus evidencing a unique form of {\it spin-orbital
coupling}. The stable configuration for the chiral $p_x+ip_y$
state evidences competition between the different orbital
components, with a
%The competing orbital components of the chiral
%$p_x+ip_y$ state {\clb are responsible} a
first-order transition from the perpendicular to the parallel
configuration occurs as the FM exchange field is increased. When
the interface is imperfect, other processes play the decisive role
in setting the easy axis in the FM.

\begin{figure}
\centering{
\includegraphics[width=0.67 \linewidth,clip=]{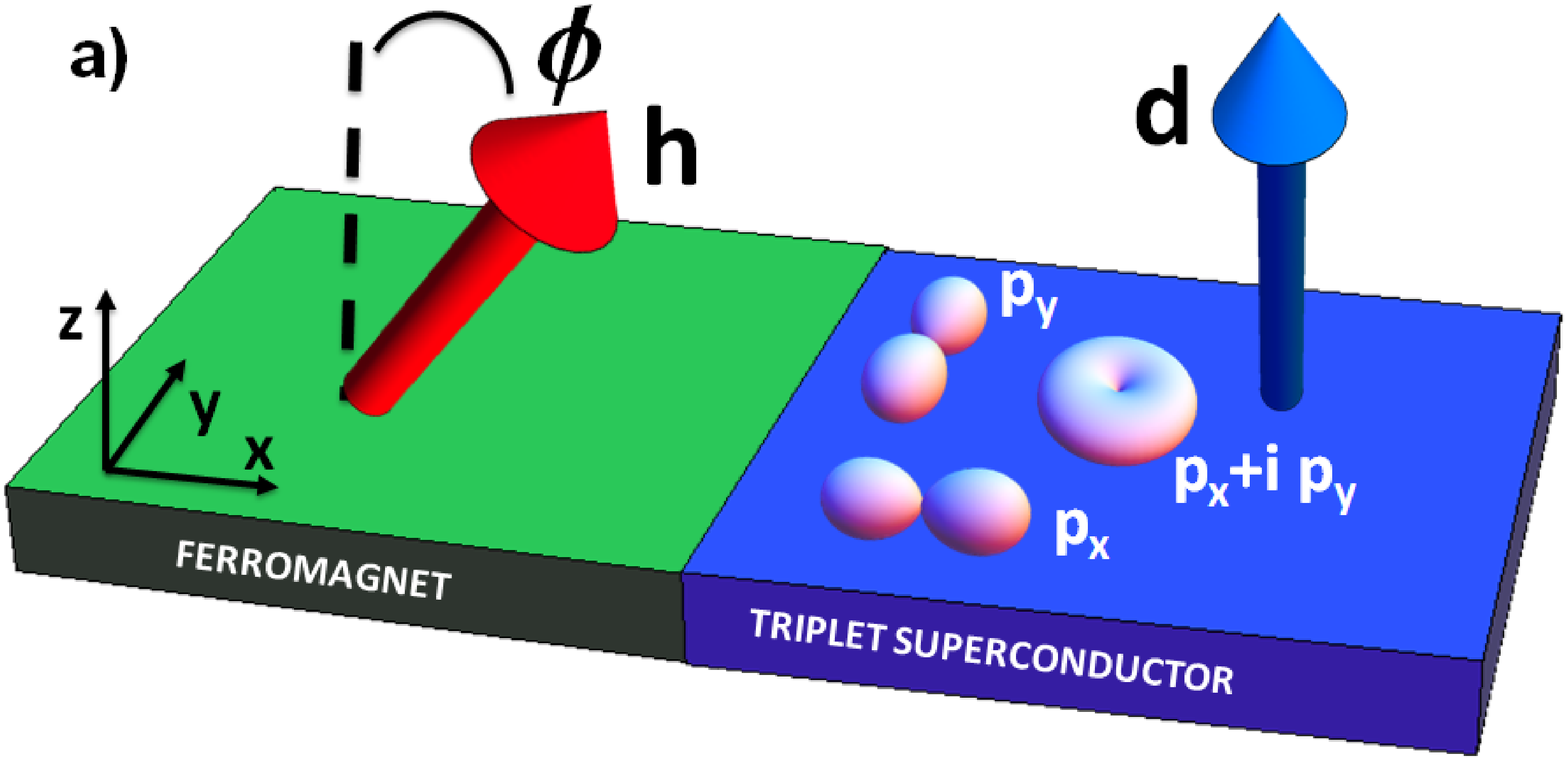}\includegraphics[width=0.33 \linewidth,clip=]{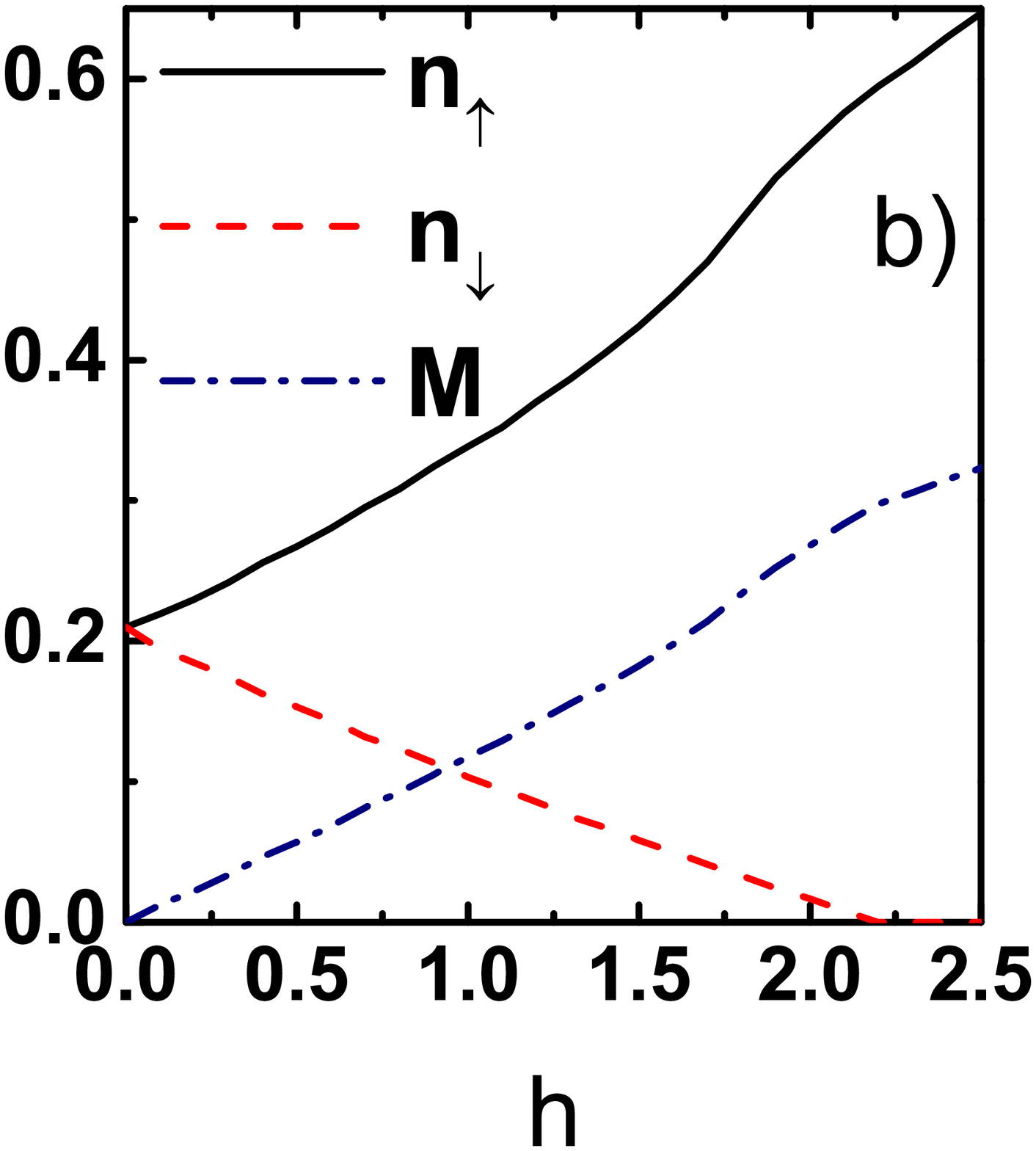}
\caption{(Color online) (a) Schematic diagram of the
two-dimensional TSC-FM junction. The FM region is located at
$x<0$, while the TSC is realized for $x>0$. The magnetization
${\bf M}$ of the FM is collinear to the exchange field ${\bf h}$
and forms an angle $\phi$ with the ${\bf d}$-vector of the TSC,
which defines the $z$ axis. We study TSC states with $p_x$, $p_y$
and $p_x+ip_y$ symmetry. (b) Evolution of the bulk FM
magnetization and the majority and minority spin concentrations as
a function of the exchange field ${\bf h}$.}
%{\clb The magnetization is collinear to the exchange field}. }
\label{fig:fig1}}
\end{figure}

{\it The model}. We examine a lattice model of the TSC-FM junction
shown in~\fig{fig:fig1}(a). The
% We study a finite-size square lattice model of
%the planar TSC-FM heterostructure shown in~\fig{fig:fig1}(a). The
lattice size is $(L+1)\times (L+1)$, with periodic boundary
conditions imposed along the direction parallel to the interface.
Indicating each site by a vector $\mathbf{i}\equiv(i_x,i_y)$, with
$i_x$ and $i_y$ being integers ranging from $-L/2$ to $L/2$, we
write the Hamiltonian
\begin{eqnarray} H&=& -\sum_{\langle \mathbf{i}
,\mathbf{j} \rangle,\,\sigma} t_{\mathbf{i},\mathbf{j}}(c^{\dagger}_{\mathbf{i}\,\sigma}
c_{\mathbf{j}\,\sigma}+\Hc) -\mu \sum_{\mathbf{i},\sigma}
n_{\mathbf{i}\sigma} \nonumber \\&& - \sum_{\langle \mathbf{i}
,\mathbf{j} \rangle \in \text{TSC}} V\left( n_{\mathbf{i}
\uparrow} n_{\mathbf{j}\downarrow}+n_{\mathbf{i}\downarrow}
n_{\mathbf{j} \uparrow} \right) - \sum_{\mathbf{i} \in \text{FM}}
\mathbf{h}\cdot\mathbf{s}_{\mathbf i}\;, \label{eq:Ham}
\end{eqnarray}
where $c_{\mathbf{i}\,\sigma}$ is the annihilation operator of an
electron with spin $\sigma$ at the site ${\mathbf{i}}$,
$n_{\mathbf{i}\,\sigma} = c^{\dagger}_{\mathbf{i}\,\sigma}
c_{\mathbf{i}\,\sigma}$ is the spin-$\sigma$ number operator, and
$\mathbf{s}_{\mathbf i} =
\sum_{s,s'}c^{\dagger}_{\mathbf{i}\,s}{\pmb{\sigma}}_{s,s'}c_{\mathbf{i}\,s'}$
is the local spin density. The lattice is divided into three
regions: the FM subsystem for $i_x<0$, the TSC subsystem for
$i_x>0$, and the interface at $i_x=0$. The chemical potential
$\mu$ is the same across the lattice. The hopping matrix elements
$t_{\mathbf{i},\mathbf{j}}=t$ everywhere, except for hopping
between the ordered subsystems where
$t_{\mathbf{i},\mathbf{j}}=t_{\text{int}} > t$ ($<t$) models an
imperfect interface with enhanced (suppressed) charge transfer
probability. All energy scales are expressed in units of $t$. A
nearest-neighbor attractive interaction $-V < 0$ is present only
on the TSC side of the junction. The order parameter of the TSC,
the so-called ${\bf d}$-vector, encodes the intrinsic spin
structure of the Cooper pairs: the ${\bf d}$-vector is defined as
${\bf d}=\frac{1}{2}(\Delta_{1}-\Delta_{-1})\hat{\bf{x}}-
\frac{i}{2}(\Delta_{1}+\Delta_{-1})\hat{\bf{y}}+\Delta_{0}\hat{\bf{z}}$,
where $\Delta_{S_z}$ is the gap for triplet pairing with $z$
component of the spin $S_z=-1, 0, 1$. In our model,~\eq{eq:Ham}, a
TSC state with ${\bf d}$-vector parallel to the $z$-axis can
stabilized at mean-field level by tuning the electron density and
pairing strength.
%The TSC order parameter is the so-called ${\bf d}$-vector, defined
%in spin space ${\bf d}=\frac{1}{2}(\Delta_{1}-\Delta_{-1})\hat{\bf{x}}- \frac{i%}{2}(\Delta_{1}+\Delta_{-1})\hat{\bf{x}}+\Delta_{0}\hat{\bf{z}}$.
%By tuning the electron density
%and the pairing strength a TSC state with
%parallel to the $z$-axis can be stabilized at mean-field level.
We consider $p_x$, $p_y$, and $p_x+ip_y$ orbital symmetries for the
pairing amplitude. The FM subsystem is modelled by the exchange
field ${\bf h}$, which forms the angle $\phi$ with respect to the
direction of the ${\bf d}$-vector. Since the TSC state is
invariant under spin rotations about the ${\bf d}$-vector, ${\bf
h}$ can be restricted to the $x$-$z$ plane, i.e. ${\bf
h}=h(\sin(\phi),0, \cos(\phi))$. The relation between the
amplitude of the magnetization ${\bf M}$ and ${\bf h}$ is shown
in~\fig{fig:fig1}(b), with ${\bf M}$ being collinear to ${\bf h}$.

We obtain a single-particle Hamiltonian $H_{MF}$ from~\eq{eq:Ham}
by decoupling the interaction term and solving self-consistently
for the mean-field amplitudes
$\Delta_{\mathbf{i}\mathbf{j}}=\langle c_{\mathbf{i}\,\uparrow}
c_{\mathbf{j}\,\downarrow} \rangle$, with the average $\langle A
\rangle$ being the thermal expectation value of the operator
$A$~\cite{supp}. We hence calculate the condensation energy
$E_\Delta$ of the TSC and the Gibbs free energy $F$ of the
junction,
\begin{eqnarray}
E_{\Delta}&=& \frac{|V|}{L^2} \sum_{\langle \mathbf{i} ,\mathbf{j}
\rangle \in \text{TSC}}
|\Delta_{\mathbf{i}\mathbf{j}}|^2\,, \\
F&=&-\frac{1}{L^2 \beta} \ln \left(\mbox{Tr}\left\{ \exp[-\beta
H_{MF}] \right\}\right)\,,
\end{eqnarray}
where $\beta=(k_B T)^{-1}$ and $k_B$ is the Boltzmann constant.
The magnetization is determined by summing over the local spin
density in the FM region, i.e. ${\bf{M}}=\frac{4}{L^2}
\sum_{i\subset FM} \langle {\bf s}_i \rangle$.
%Both $F$ and $E_{\Delta}$ depend
%on the exchange field and the angle $\phi$ through the self-consistent pairing
%amplitude and magnetization.
The results presented here were obtained using $L=120$; a larger
lattice does not qualitatively change our conclusions.

In our analysis of the TSC-FM junction we first aim to understand
how the pairing potential changes with the magnetization
orientation. For this it is convenient to assume that the angle
$\phi$ is {\it fixed}. The observed changes in the pairing
potential as a function of $\phi$ then motivates us to treat
$\phi$ as a {\it variational parameter}, and to seek the most
stable magnetic configuration. These individual steps have
physical relevance: the former models the case where the
magnetization in the FM is strongly pinned by anisotropy or an
external field, whereas the latter corresponds to the limit of an
isotropic FM where the TSC acts as the unique source of spin
symmetry breaking.
%{\clb The
%strategy we follow for the analysis is to first understand how the
%pairing potential changes with the magnetization orientation,
%assuming a fixed angle $\phi$ for the exchange field. Secondly, we
%determine the most stable magnetic configuration, requiring us to
%treat $\phi$ as a variational parameter. These two steps are
%physically relevant for the cases when the magnetization in the
%ferromagnet is strongly pinned by anisotropy or an external field
%(i.e. a fixed $\phi$) and an isotropic ferromagnet where the
%magnetization has no preferential orientation and the spin-triplet
%superconductor acts as the unique source of spin symmetry breaking
%(i.e. a variable $\phi$).}

\begin{figure}
\centering{
\includegraphics[width=\columnwidth,clip]{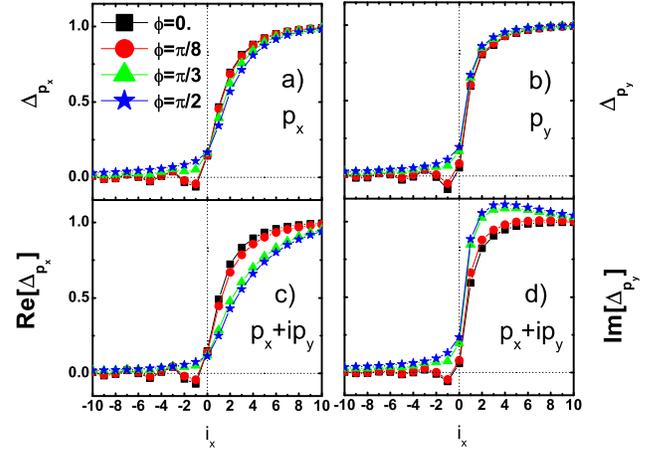}
\caption{(Color online) Zero temperature pairing amplitude scaled
to its bulk value as a function of the distance $i_x$ from the
interface for $h=1.5$, $t_{\text{int}}=1$, and several different
angles $\phi$ of the ${\bf
  d}$-${\bf M}$
misalignment. The spin-triplet orbital symmetry is of (a) $p_x$,
(b) $p_y$ , and chiral type with (c) a real $p_x$ and (d) an
imaginary $p_y$ component, respectively. } \label{fig:fig2}}
\end{figure}

{\it Pairing amplitude.} In Fig. \ref{fig:fig2} we present the
pairing amplitude profile near the interface for $h=1.5$,
$t_{\text{int}}=1$, and several different values of
$0\leq\phi\leq\frac{\pi}{2}$. This is determined by minimizing the
Gibbs energy functional with respect to the pairing amplitudes at
fixed angle~\cite{supp}. Distinct trends are evident both in the
FM and TSC sides of the junction as the exchange field is rotated
from a parallel ($\phi=0$) to a perpendicular
($\phi=\frac{\pi}{2}$) orientation with respect to the ${\bf
d}$-vector. Independent of the orbital symmetry, the proximity
effect in the FM smoothly evolves from a monotonous decay at $\phi
= \frac{\pi}{2}$ to a damped oscillating behaviour at $\phi=0$.
The oscillating behaviour is similar to that observed in an SSC-FM
junction~\cite{SCFMreviews}, and is also due to pair breaking,
specifically the {\it spin-spin} coupling between the
$z$-component of the exchange field and the in-plane spin of the
triplet Cooper pairs.

In contrast, the pairing amplitude on the TSC side of the
interface strongly depends upon both the angle $\phi$ and the
\emph{orbital} symmetry of the TSC. For a TSC with $p_x$ orbital
symmetry the pairing amplitude near the interface is reduced as
the exchange field is tilted from parallel to perpendicular with
respect to the ${\bf d}$-vector[see~\fig{fig:fig2}(a)]; the
opposite behaviour is observed for a $p_y$ TSC, although the
effect is less pronounced [see~\fig{fig:fig2}(b)]. The chiral
$p_x+ip_y$ TSC evidences both trends: decreasing $\phi$ from
$\frac{\pi}{2}$ to $0$ enhances the real ($p_x$) part of the gap
[\fig{fig:fig2}(c)], but suppresses the imaginary ($p_y$) part
[\fig{fig:fig2}(d)]. Competition between the two gap components
enhances their variation with $\phi$ compared to the time-reversal
symmetric states.

The pair-breaking due the spin-spin coupling cannot explain the
different $\phi$-dependence of the $p_y$ and $p_x$ gap profiles.
This instead originates from the spin-flip reflection of triplet
Cooper pairs at the interface with the FM, which is crucial for
the spin Josephson effect~\cite{Brydon2009,Brydon2011}. In such a
scattering process, an incident Cooper pair with spin $\sigma$
mutually perpendicular to ${\bf d}$ and ${\bf M}$ acquires the
spin- and orbital-dependent phase shift $\pi - 2\sigma\phi +
\Delta\theta$: the first two terms are due to the spin-flip, while
the last is due to the phase change of the TSC gap upon specular
reflection. Here $\Delta\theta=\pi$ ($0$) for the $p_x$ ($p_y$)
state, while $\Delta\theta$ depends on the angle of incidence for
the $p_x+ip_y$ gap. It is well known that the gap is suppressed at
interfaces where reflected Cooper pairs undergo a non-trivial
phase shift~\cite{gapsuppress}; in the TSC-FM junction we hence
maximize the gap at the interface by choosing $\phi$ so that the
spin-flip reflected Cooper pairs have a $2\pi{n}$ phase shift. Due
to the different orbital phase shifts $\Delta\theta$, this occurs
at $\phi=0$ ($\frac{\pi}{2}$) for the $p_x$ ($p_y$) pairing
amplitude, in agreement with~\fig{fig:fig2}. This interplay of
spin and orbital degrees of freedom manifests an unconventional
type of spin-orbital coupling at the TSC-FM interface.
\begin{figure}
\centering{
\includegraphics[width=\columnwidth,clip=]{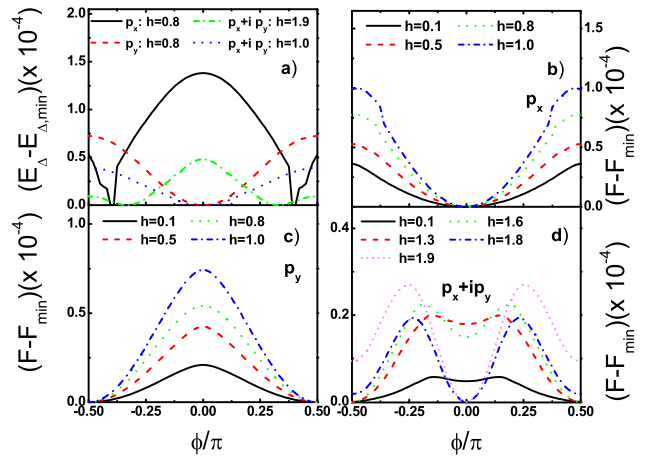}
\caption{(Color online) (a) Dependence of the condensation energy
$E_\Delta$ on the angle $\phi$. (b-d) Dependence of the Gibbs
energy $F$ on $\phi$ for various fixed $h$ and for $p_x$, $p_y$
and $p_x+ip_y$ orbital symmetries of the TSC. $E_{\Delta,{\text
{min}}}$ and $F_{\text{min}}$ are the minimum amplitudes of the
related energies. All panels are for $t_{\text{int}}=1$ and
temperature $k_B T=0.05$.} \label{fig:fig3}}
\end{figure}

{\it Stable moment orientation.} The spin-spin and spin-orbital
coupling effects give $\phi$-dependent contributions to the Gibbs
free energy $F$ of the junction, e.g., by modifying the local
density of states in the FM and the condensation energy $E_\Delta$
in the TSC, respectively. The energetically-favored moment
orientation is found directly from $F$, while the relevance of the
spin-orbital coupling can be deduced from $E_{\Delta}$.
In~\fig{fig:fig3} we present the behavior of the Gibbs energy $F$
and the condensation energy $E_\Delta$ as a function of $\phi$,
where these quantities are evaluated for the pairing amplitudes
that minimize $F$ at the given angle~\cite{supp}.

In~\fig{fig:fig3}(a) we plot $E_\Delta$ as a function of $\phi$
for several typical cases and a perfect interface
($t_{\text{int}}=1$). As expected, the condensation energy for the
$p_x$ and $p_y$ TSCs is indeed maximized for the exchange field
orientation which maximizes the gap amplitude. The $p_x+ip_y$ case
is more complicated, since here the $p_x$ and $p_y$ gap components
show opposite dependence upon $\phi$. We find that the maximum in
$E_{\Delta}$ shifts from $\phi=\frac{\pi}{2}$ to $\phi=0$ with
increasing exchange field strength. That is, for a weak FM the
$p_y$ component dominates the physics, while at strong
polarizations the $p_x$ gap is most important.

\begin{figure}%[h!]
\centering{
\includegraphics[width=\columnwidth,clip=]{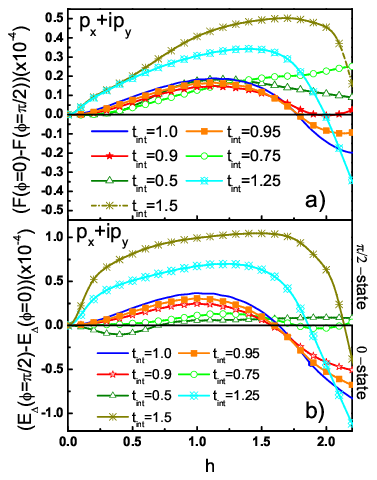}
\includegraphics[width=\columnwidth,clip=]{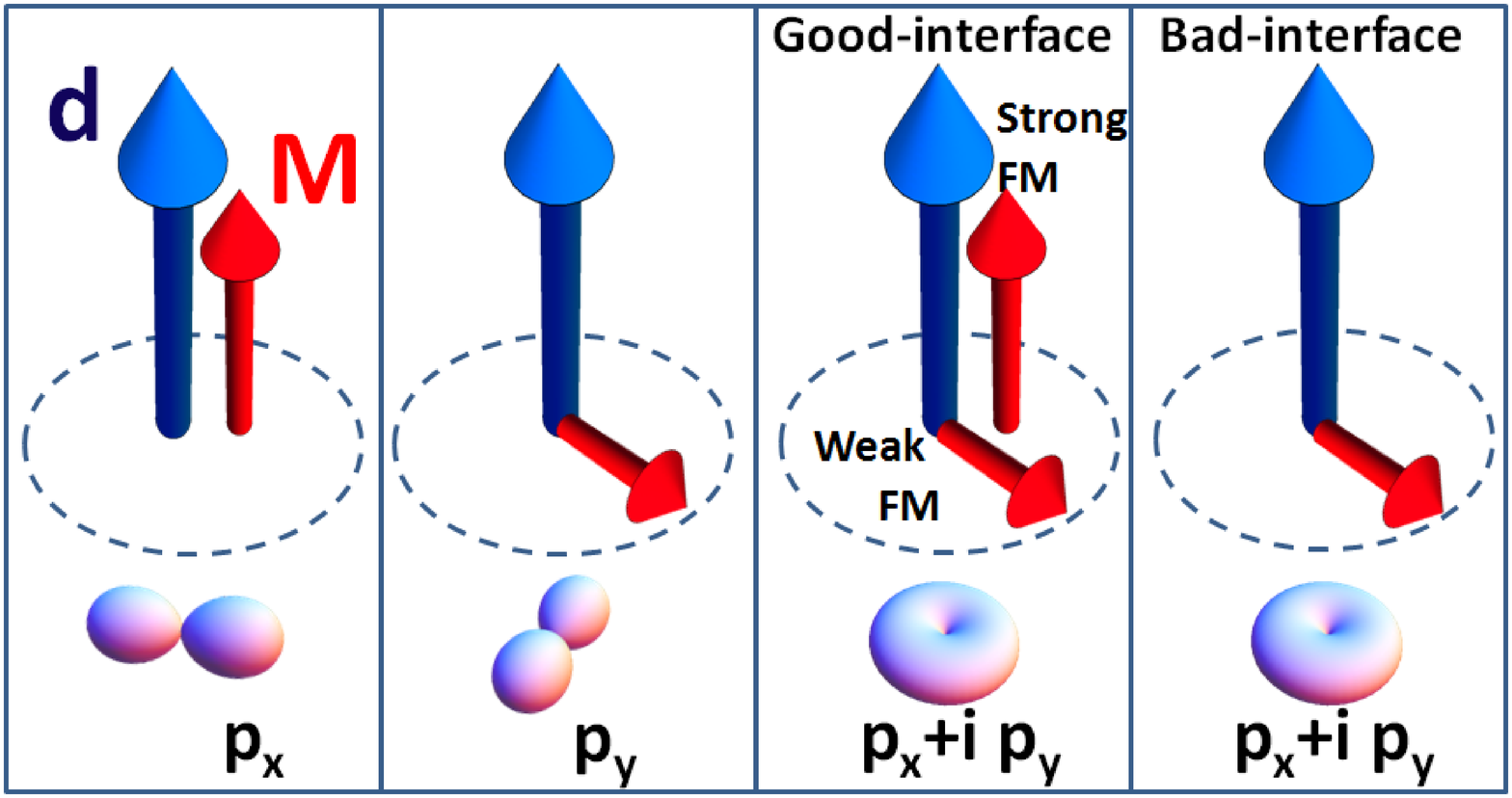}
\caption{(Color online) (a) Gibbs energy and (b) order parameter
energy difference between the parallel ($\phi=0$) and
perpendicular ($\phi=\frac{\pi}{2}$) configurations of the moment
as a function of the exchange field strength $h$ and
$t_{\text{int}}$, at temperature $k_B T=0.03$ for the $p_x+ip_y$
TSC. Bottom panel: sketch of the most favorable magnetic (small
red arrow) configurations with respect to the orbital symmetry and
the ${\bf d}$-vector (large blue arrow) of the TSC in terms of the
interface transparency character as well.} \label{fig:fig4}}
\end{figure}

The minimum of the Gibbs free energy $F$ fixes the stable moment
orientation. In~\fig{fig:fig3}(b)-(d) we plot $F$ as a function of
$\phi$ for the three orbital symmetries at $t_{\text{int}}=1$. For
the $p_x$ orbital symmetry, the profile exhibits a single minimum
at $\phi=0$ and a maximum at $\phi=\frac{\pi}{2}$, and \emph{vice
versa} for the $p_y$ TSC. The stable magnetic orientation is
therefore parallel (perpendicular) to the ${\bf d}$-vector if the
antinodes of the $p$-wave TSC gap are perpendicular (parallel) to
the interface.
%Although here this is primarily controlled by the
%spin-orbital coupling via the condensation energy, interface
%electronic reconstruction is expected to favour the same
%magnetization configurations~\cite{Brydon2011}.
Our conclusions are robust to changing $t_{\text{int}}$ as shown
in the supplemental material.

The Gibbs free energy for the $p_x+ip_y$ junction has minima at
both $\phi=0$ and $\phi=\frac{\pi}{2}$, which is not anticipated
from the condensation energy. As shown in~\fig{fig:fig3}(d)
and~\fig{fig:fig4}(a), at $t_{\text{int}}=1$ the global minimum
shifts from $\phi=\frac{\pi}{2}$ (perpendicular) to $\phi=0$
(parallel) with increasing exchange field strength. This occurs at
the critical field strength $h_{cr,1}\approx1.74$, which is a
little higher than if we considered only the condensation energy
[see~\fig{fig:fig4}(b)]. Further increasing $h$ into the extreme
half-metal regime, we find that the $\phi=\frac{\pi}{2}$ state
reappears above $h_{cr,2}$ (not shown). The two critical fields
merge together as the temperature is increased, so that only the
$\phi=\frac{\pi}{2}$ is stable sufficiently close to $T_c$. While
the $t_{\text{int}}>1$ results are qualitatively similar, reducing
$t_{\text{int}}$ entirely suppresses the $\phi=0$ state. Here we
also observe a decoupling between the condensation energy gain and
the Gibbs free energy, e.g. for $t_{\text{int}} \lesssim 0.8$ the
gain in condensation energy in the low-field regime favors a
$\phi=0$ state, while the Gibbs free energy shows that the
$\phi=\frac{\pi}{2}$ state is stable. Since the contribution to
the free energy in the TSC region can be ascribed to $E_{\Delta}$,
any inconsistency between the location of the minimum in $F$ and
$E_{\Delta}$ must be due to the changes in the energy spectrum at
the interface and in the FM, which are only included in the
former. We expect that the modification of the energy spectrum
will depend rather strongly upon the interface hopping
$t_{\text{int}}$, and indeed in \fig{fig:fig4} we observe that the
condensation energy tends to overstate the stability of the
$\phi=0$ state.
%When monitoring the role of the hopping at the interface we can
%deduce that, if $F$ is minimized by a configuration that is not
%the minimum for $E_{\Delta}$, it is the modification of the energy
%spectrum close to the interface, being dependent on the magnetic
%orientation of the ferromagnet, the driving term in settling the
%magnetic configuration.
We conclude that for a sufficiently imperfect interface the
magnetization orientation is controlled by other processes, such
as the change of the spectrum at the interface ~\cite{Brydon2011}
or the proximity effect.

%Our results imply a close connection between the spin-orbital
%coupling and the spin Josephson effect in the TSC-FM
%junction~\cite{Brydon2009,Brydon2011}, at least for the $p_x$ and
%$p_y$ states. Specifically, the Josephson spin current is
%generated by a $\phi$-dependent Helmholtz free energy; for a
%non-self-consistent model of a thin FM layer on a bulk TSC, the
%free energy due to the interface electronic reconstruction has
%similar form to that obtained here and gives the same  stable
%magnetization configurations.
%for the
%$p_x$ and $p_y$ states~\cite{Brydon2011}.

{\it Experimental considerations.} The apparently small energy
difference between the $\phi=0$ and $\frac{\pi}{2}$ states shown
in~\fig{fig:fig4} results from averaging what is essentially an
interface effect over the entire lattice; the energy gain per
interface unit cell is $L$ times larger, and gives an anisotropy
energy on the order of $\sim 0.01 k_BT_c$ for the microscopic
parameters chosen here.
%In the case of a thin FM
%layer of width a few unit cells, this is comparable to the free energy
%change due to electronic reconstruction within the FM
%itself~\cite{Brydon2011}.
In the case of a thin FM layer, the magnetic anisotropy induced by
the coupling to the TSC could be observed by ferromagnetic
resonance (FMR) measurements: for an exchange field $h=0.5$ in the
FM, and choosing Sr$_2$RuO$_4$ ($T_c=1.5K$) for the bulk TSC, we
estimate a precession frequency of $\sim5\cos(\phi)$~GHz. Since
there is no spin-orbital coupling at SSC-FM interfaces, the
observation of this precession would strongly indicate a TSC state
in the superconductor. For a thicker layer, the TSC can modify the
magnetization profile near the interface, effectively creating a
spin-active boundary layer~\cite{Terrade2012}. This may
qualitatively alter the proximity effect and the electronic
transport properties of the junction~\cite{spinactive}.

{\it Summary.} In this Letter we have studied the interplay
between orbital and spin degrees of freedom in a TSC-FM
heterostructure. The orbital pairing state in the bulk TSC plays a
critical role in fixing the stable orientation of the
magnetization in the FM, which is summarized by the sketch
in~\fig{fig:fig4}. For the time-reversal symmetric gaps the easy
axis in the FM originates from the maximization of the TSC's
condensation energy.
%For a TSC with a
%single gap component, an FM moment parallel (perpendicular) to the
%TSC ${\bf d}$-vector maximizes the magnitude of a $p$-wave order
%parameter with antinodes perpendicular (parallel) to the
%interface, and the resulting gain in condensation energy induces
%the corresponding easy axis in the FM.
On the other hand, the orbital frustration of the condensation
energy in a chiral TSC leads to a magnetic configuration with a
first-order transition between the perpendicular and parallel
configurations as a function of the exchange field. Spin-dependent
electronic reconstruction at an imperfect interface can compensate
the condensation energy gain. We argue that the induced anisotropy
axis in the FM could be observed in FMR measurements, and can act
as a test of the orbital and spin pairing state of the TSC.

{\it Acknowledgements.} The authors thank M. Sigrist and C. Timm
for useful discussions. This research was supported by the EU
-FP7/2007-2013 under grant agreement N. 264098 - MAMA.

\end{document}

\title{Supplementary material for ``Spin-orbital coupling in a triplet superconductor--ferromagnet junction''}

\author{Paola Gentile}
\affiliation{SPIN-CNR, I-84084 Fisciano (Salerno), Italy}
\affiliation{Dipartimento di Fisica ``E. R. Caianiello'',
Universit\`{a} di
  Salerno, I-84084 Fisciano (Salerno), Italy }
\author{Mario Cuoco}
\affiliation{SPIN-CNR, I-84084 Fisciano (Salerno), Italy}
\affiliation{Dipartimento di Fisica ``E. R. Caianiello'',
Universit\`{a} di
  Salerno, I-84084 Fisciano (Salerno), Italy }
\author{Alfonso Romano}
\affiliation{SPIN-CNR, I-84084 Fisciano (Salerno), Italy}
\affiliation{Dipartimento di Fisica ``E. R. Caianiello'',
Universit\`{a} di
  Salerno, I-84084 Fisciano (Salerno), Italy }
\author{Canio Noce}
\affiliation{SPIN-CNR, I-84084 Fisciano (Salerno), Italy}
\affiliation{Dipartimento di Fisica ``E. R. Caianiello'',
Universit\`{a} di
  Salerno, I-84084 Fisciano (Salerno), Italy }
\author{Dirk Manske}
\affiliation{Max-Planck-Institut f\"{u}r Festk\"{o}rperforschung,
  Heisenbergstr. 1, D-70569 Stuttgart, Germany}
\author{P. M. R. Brydon}
\affiliation{Institut f\"{u}r Theoretische Physik, Technische
Universit\"{a}t Dresden, D-01062 Dresden, Germany}

\begin{abstract}
In this supplementary material we present the analysis of the
optimal ferromagnetic orientation due to the interface
transparency for the case of $p_x$ and $p_y$ spin triplet
superconductors. Furthermore, we provide extra details about the
scheme adopted for determining the self-consistent profiles of the
pairing amplitude and of the magnetization.
\end{abstract}

\maketitle

\begin{figure}[btp]
\centering{
\epsfig{file=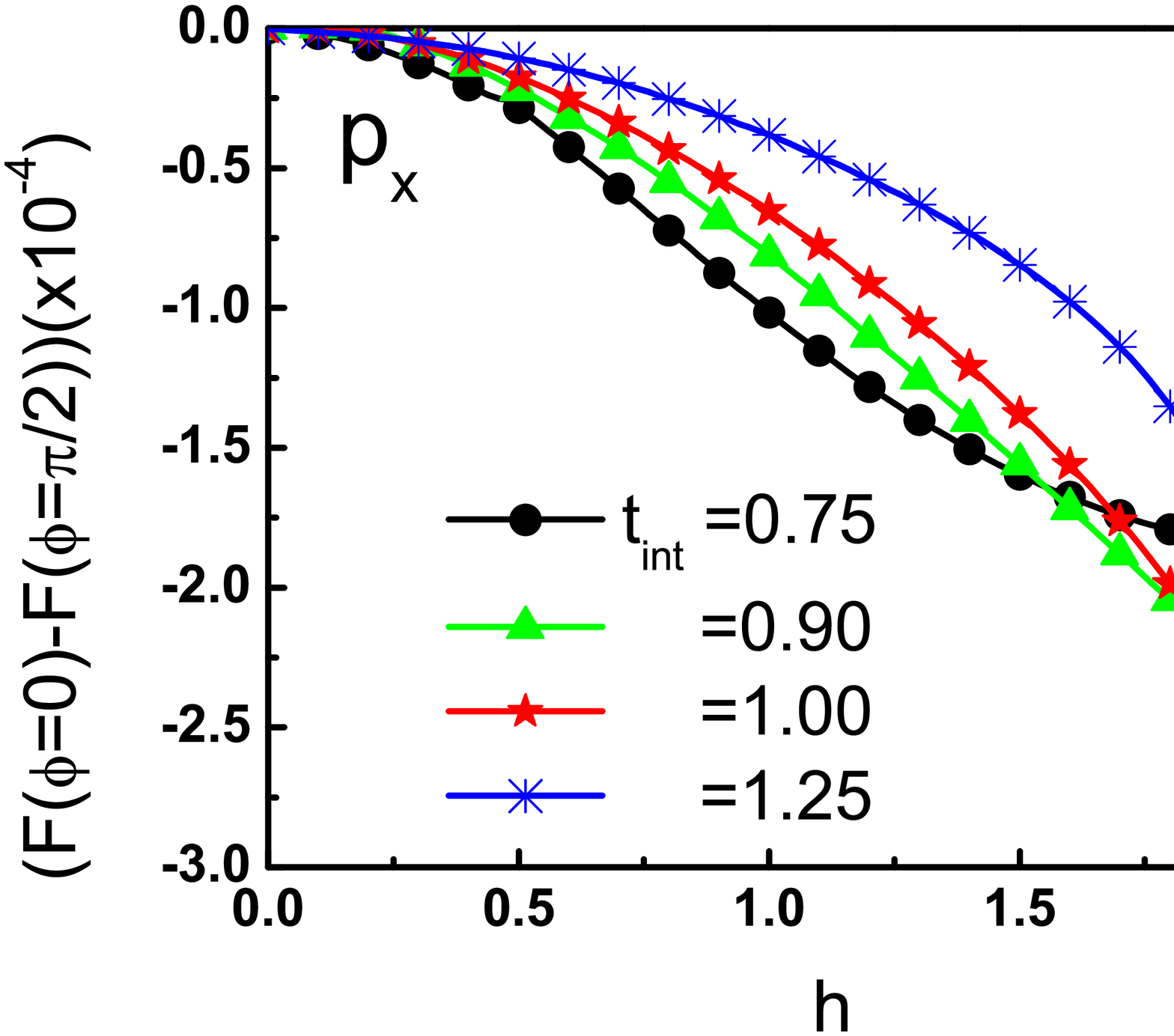,width=0.90\linewidth} \\
\caption{(color online) Gibbs free energy difference between the
parallel ($\phi=0$) and perpendicular ($\phi=\frac{\pi}{2}$)
configurations of the ferromagnetic moment with respect to the
${\bf d}$-vector as a function of the exchange field strength $h$
at a representative temperature $k_{B} T=0.03$ for the $p_x$ TSC.
} \label{fig:fig-px}}
\end{figure}

\section{Mean-field solution}

To numerically solve the model Hamiltonian of Eq. 1 we decouple
the interaction term in the Hartree-Fock approximation by
introducing the pairing amplitude on a bond,
$\Delta_{\mathbf{i}\mathbf{j}}=\langle c_{\mathbf{i}\,\uparrow}
c_{\mathbf{j}\,\downarrow} \rangle$, so that
\begin{eqnarray*}
V n_{\mathbf{i}\uparrow} n_{\mathbf{j}\downarrow}\simeq
(\Delta_{\mathbf{i}\mathbf{j}}
c^{\dagger}_{\mathbf{j}\,\downarrow}
c^{\dagger}_{\mathbf{i}\,\uparrow}+\Delta^{*}_{\mathbf{i}\mathbf{j}}
c_{\mathbf{i}\,\uparrow}
c_{\mathbf{j}\,\downarrow}-|\Delta_{\mathbf{i}\mathbf{j}}|^2) \,,
\end{eqnarray*}
where the average $\langle A \rangle$ indicates the thermal
expectation value of the operator $A$. After the decoupling the
Hamiltonian reduces to a bilinear form, $H_{MF}$, which can be
diagonalized by means of a Bogoliubov-de Gennes transformation.
The pairing amplitudes on each bond can yield both spin singlet
($S$) and triplet ($T$) components,
\begin{eqnarray*}
\Delta^{S,T}=(\Delta_{\mathbf{i}\mathbf{j}}\pm\Delta_{\mathbf{j}\mathbf{i}})/2,
\end{eqnarray*}
which in turn define superconducting order parameters with $s$-
or $p$-wave symmetry, i.e.
\begin{eqnarray*}
\Delta_{s}(\mathbf{i})=\frac{1}{4}(\Delta^{S}_{\mathbf{i},\mathbf{i}+\text{\^{x}}}+\Delta^{S}_{\mathbf{i},\mathbf{i}-\text{\^{x}}}+
\Delta^{S}_{\mathbf{i},\mathbf{i}+\text{\^{y}}}+\Delta^{S}_{\mathbf{i},\mathbf{i}-\text{\^{y}}})
\end{eqnarray*}
and
\begin{eqnarray*}
\Delta_{p_{x(y)}}(\mathbf{i})=\frac{1}{2}(\Delta^{T}_{\mathbf{i},\mathbf{i}+\text{\^{x}}(\text{\^{y}})}-\Delta^{T}_{\mathbf{i},\mathbf{i}-\text{\^{x}}(\text{\^{y}})}),
\end{eqnarray*}
with $\text{\^{x}} (\text{\^{y}})$ being the unit vectors
connecting the nearest neighbor sites along the direction
perpendicular (parallel) to the interface.

Although spin singlet superconductivity can be generated by the
nearest-neighbour pairing interaction, this is not stable in the
parameter range where the analysis has been performed
~\cite{Cuoco2008}. The stable pairing state in our model is
controlled by the electron density, and $p$-wave triplet solutions
are realized in the region between half-filling, $\mu\simeq0$, and
high (low) density ($|\mu|\simeq 2.5$)~\cite{Kuboki01}. We choose
the chemical potential to be {$|\mu|= 1.8$, in the window of
stability. For this set of parameters the bulk TSC critical
temperature is $k_BT_c \simeq 0.18$.} Anisotropic pairing
interactions with $V_{x}=-V$, $V_y=0$ and {\it vice versa} are
used to model the $p_x$ and $p_y$ states, respectively.

\subsection{Computational algorithm}

We briefly sketch the computational procedure used to determine
    the solutions for the optimal pairing amplitude and the stable
    magnetization direction.
\begin{enumerate}
\item The decoupling of the quartic term on the superconducting side
  introduces a set of
  $2L^2$ variational parameters $\Lambda_{\mathbf{i}} = \{
\Delta_{\mathbf{i},\mathbf{i}+\text{\^{x}}},
\Delta_{\mathbf{i},\mathbf{i}-\text{\^{x}}},\Delta_{\mathbf{i},\mathbf{i}+\text{\^{y}}}
,\Delta_{\mathbf{i},\mathbf{i}-\text{\^{y}}} \}$, with
$\text{\^{x}} (\text{\^{y}})$ being the unit vectors connecting
the nearest neighbor sites along the direction perpendicular
(parallel) to the interface. In the ferromagnet (FM) the angle
$\phi$ can be treated as a variational parameter, although it does
not follow from decoupling an interaction term.  Due to the
translational invariance along the $y$ direction, the number of
variational parameters can be further reduced to $2L$. The
mean-field Hamiltonian $H_{MF}$ is, then, a function of $2L+1$
variational parameters, $H_{MF}(\Lambda_{\bf i},\phi)$.

\item For a given set of variational $\{\Lambda_{\bf i},\phi\}$, and fixed
microscopic parameters, i.e. $\{t,h,V,\mu\}$, the spectrum of the
bilinear mean-field Hamiltonian is obtained by standard
diagonalization routines.

\item At finite temperature the functional to be used for determining
the variational parameters is the Gibbs energy. This is directly
computed from the spectrum of the mean-field Hamiltonian by
performing the trace over all the eigenstates as
\begin{eqnarray*}
F(\Lambda_{\bf i},\phi)=-\frac{1}{L_x L_y \beta} \ln
\left(\mbox{Tr}\left\{ \exp[-\beta H_{MF}] \right\}\right)\,.
\end{eqnarray*}

\item The variational parameters are determined by solving the
coupled set of gap equations
\begin{eqnarray*}
&&\frac{\partial F (\Lambda_{\bf i},\phi)}{\partial \Lambda_{\bf i}}=0 \\
&& \frac{\partial F(\Lambda_{\bf i},\phi)}{\partial \phi}=0
\end{eqnarray*}
It is convenient to perform this in two steps:
\begin{enumerate}
\item we first minimize the Gibbs energy for a fixed angle $\phi$ to obtain the
  pairing amplitudes and search for the solutions $\bar{\Lambda}_{\bf
    i}(\phi)$. The pairing amplitudes are calculated iteratively until the
  difference between successive iterations is smaller than some desired
  accuracy~\cite{Cuoco2008}. From $\bar{\Lambda}_{\bf
    i}(\phi)$ we obtain the pairing amplitude profiles at different
    angles
  $\phi$ seen in Figure 2 of the letter.
\item We evaluate $\bar{\Lambda}_{\bf
    i}(\phi)$ for many angles $-\pi/2\leq \phi \leq \pi/2$ and hence obtain
  the function
\begin{equation*}
F(\phi) = F(\bar{\Lambda}_{\bf i}(\phi),\phi) = \min_{\Lambda_{\bf
    i}}F(\Lambda_{\bf i},\phi) \,,
\end{equation*}
i.e. the minimum Gibbs energy at any given angle $\phi$. Searching
for the minimum of $F(\phi)$ gives us the overall minimum Gibbs
energy, and hence determines the stable orientation of the
magnetic moment. Typical curves of $F(\phi)$ are plotted in Figure
3 of the Letter.
\end{enumerate}
\end{enumerate}

\section{Effect of imperfect interface: $p_x$ and $p_y
$ orbital symmetry}

We discuss here the robustness of the stable moment configuration
for the $p_x$ and $p_y$ TSCs in the various regimes of interface
charge transfer. For all considered $t_{\text{int}}$ we find that
the minima of the Gibbs free energy only occur at $\phi=0$ and
$\phi=\frac{\pi}{2}$ and that the stability for the $p_x$ and
$p_y$ spin triplet superconductors is only quantitatively affected
by departures from a perfect interface.

\begin{figure}[btp]
\centering{
\epsfig{file=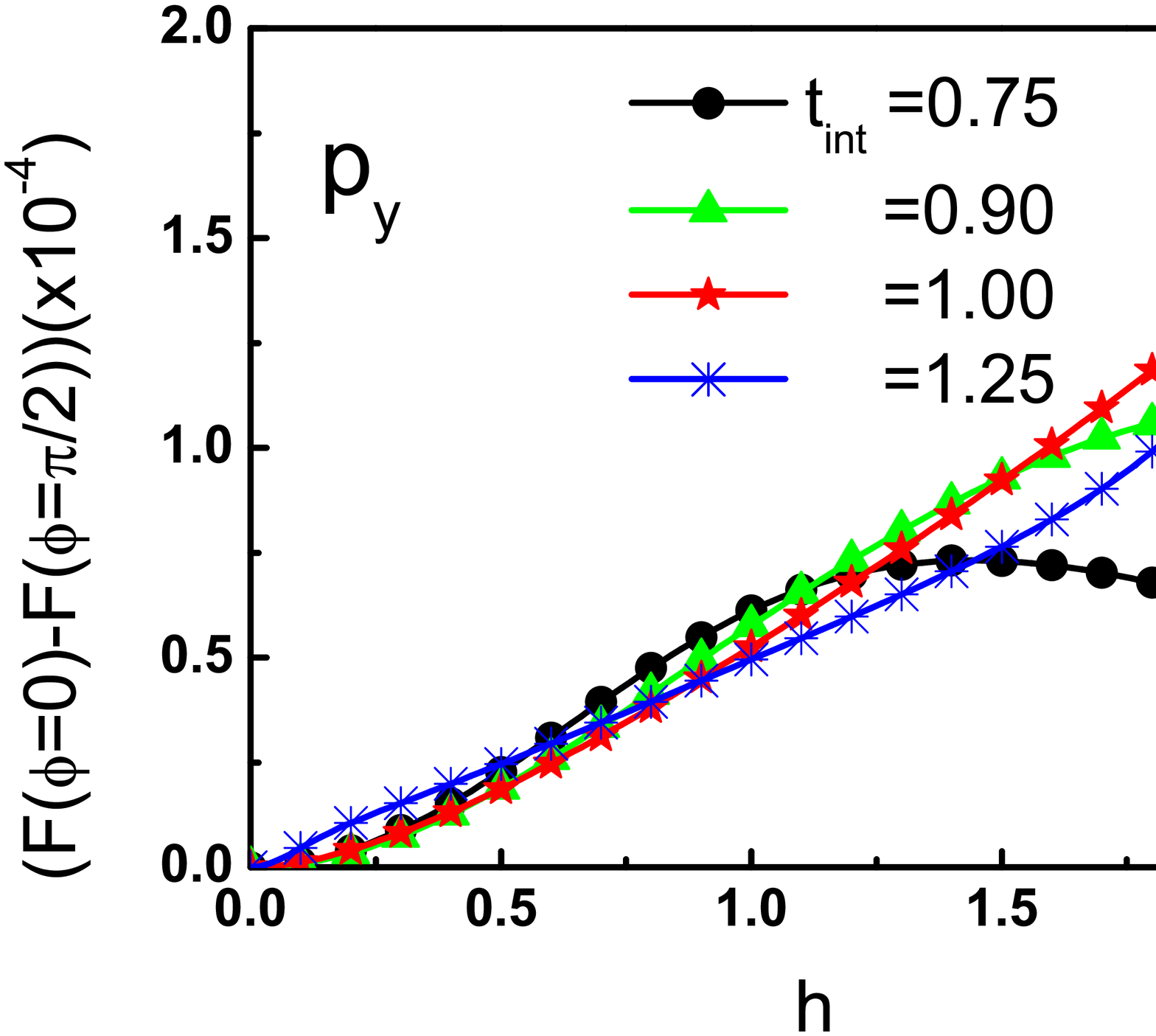,width=0.90\linewidth} \\
\caption{(color online) Same as in Fig. \ref{fig:fig-px} but for
the $p_y$ TSC.} \label{fig:fig-py}}
\end{figure}

In Figs. \ref{fig:fig-px} and \ref{fig:fig-py} we report the
energy difference between the parallel ($\phi=0$) and
perpendicular ($\phi=\frac{\pi}{2}$) orientation of the
magnetization with respect to the ${\bf d}$-vector for the Gibbs
potential considering the $p_x$ and $p_y$ cases respectively. As
one can notice, the change of the hopping amplitude at the
interface does not affect the most stable magnetic configuration
for the ferromagnet. The Gibbs potential for the $\phi=0$
($\phi=\frac{\pi}{2}$) is always the minimum for the $p_x$ ($p_y$)
TSCs at any value of the exchange field in the ferromagnet and for
both the regimes of bad and good transparency. Furthermore, the
inspection of the energy associated with the order parameter
within the superconducting side of the junction demonstrates that
for the $p_x$ symmetry the pairing amplitude at $\phi=0$ is
greater than that at $\phi=\frac{\pi}{2}$ and {\it vice versa} for
the $p_y$ TSC. Hence, the gain in the condensation energy is still
the main driving mechanism for the stability of the magnetic
orientation in the ferromagnet.

Concerning some specific features of the evolution of the Gibbs
energy for the $p_x$ TSC, we observe that the difference between
the $0$ and $\frac{\pi}{2}$ configuration (Fig.~\ref{fig:fig-px})
reveals that for weak and intermediate strength of the ferromagnet
the stability of the $0$-state is enhanced by a reduced interface
hopping. Approaching the half-metal regime (i.e. $h\simeq 2$), the
overall gain in the Gibbs energy for the $0$-configuration
decreases. Interestingly, an enhanced interface hopping
$t_{int}>1$ is always unfavorable for the $0$-configuration. This
trend is also reflected in the order parameter energy $E_{\Delta}$
where the magnetic configuration parallel to the ${\bf d}$-vector
is more detrimental for the pairing amplitude when the charge
transfer at the interface is greater that in the interior of the
junction.

For the $p_y$ symmetry, as shown in Fig. \ref{fig:fig-py}, the
difference of the Gibbs energy between the $0$ and $\frac{\pi}{2}$
configurations shows no significant changes occurring in the
regime of weak or intermediate strength of the ferromagnet either
if the hopping is reduced or increased with respect to the ideal
limit. Approaching the half-metal regime the stability of the
$\frac{\pi}{2}$-configuration is reduced when the charge transfer
at the interface diminishes. A condition for a more robust
stability of the in-plane magnetization is obtained only close to
the half-metal when the hopping is greater than that in the inner
side of the junction.

